\documentclass[useAMS,usenatbib]{mn2e}

\usepackage{amssymb}
\usepackage{graphicx}
\usepackage{times}

\def\mj{$\,${\rm M}$_{\rm J}\,$}
\def\me{$\,{\rm M}_{\oplus}\,$}
\def\h2o{H$_2$O}
\def\sio2{SiO$_2$}
\def\gc3{g\,cm$^{-3}$}

\def\threequarters{\frac{3}{4}}
\def\onequarter{\frac{1}{4}}
\def\onehalf{\frac{1}{2}}

\title[The Effect of Composition on the Evolution of Giant and Intermediate-Mass Planets]{The Effect of Composition on the Evolution of Giant and Intermediate-Mass Planets}
\author[A.~Vazan, A.~Kovetz, M.~Podolak, R.~Helled]{A. Vazan$^{1}$\thanks{E-mail:
allonava@post.tau.ac.il}, A. Kovetz$^{1,2}$, M. Podolak$^{1}$, R. Helled$^{1}$\\
$^{1}$Department of Geophysics, Atmospheric, and Planetary Sciences, Tel-Aviv University, Israel\\
$^{2}$School of Physics and Astronomy, Tel-Aviv University, Israel }
\begin{document}

\date{Accepted 2013 July 5. Received 2013 July 4; in original form 2013 April 18}

\maketitle

\label{firstpage}

\begin{abstract}

We model the evolution of planets with various masses and compositions.  We investigate the effects of the composition and its depth dependence on the long-term evolution of the planets. The effects of opacity and stellar irradiation are also considered. It is shown that the change in radius due to various compositions can be significantly smaller than the change in radius caused by the opacity. Irradiation also affects the planetary contraction but is found to be less important than the opacity effects. We suggest that the mass-radius relationship used for characterization of observed extrasolar planets should be taken with great caution since different  physical conditions can result in very different mass-radius relationships. 

\end{abstract}


\section{Introduction}
In order to characterize the many hundreds of exoplanets that have been discovered, it is necessary to determine their internal structure and composition. Detailed structural models such as those computed for the planets of our solar system are not yet possible due to lack of the necessary observationally-measured parameters.  However, for many exoplanets masses and radii are available, and a number of studies have looked at the mass-radius relationship expected for such objects. 

Some studies arrive at an empirical mass-radius correlation based on observational data \citep{weiss13}, and others obtain the mass-radius relation from theoretical evolution / interior models. \cite{marley07} studied the early evolution of gas giants and \cite{mordasini12} examined the mass-radius relationship in the context of formation scenarios, but these studies relied on one particular choice of opacity.  \cite{fortney07} computed model radii of various compositions and planetary masses by using simplified equation of state (EOS). \cite{baraffe08} modeled gas giants and super-Earths using EOS for water and rocky material, including SESAME and ANEOS. It was shown that for certain conditions the radius-mass relation can be substantially affected by the composition of the high-Z material, its mass fraction, its distribution within the planet, and the EOS used to describe it.  Here too, the assumed opacity was the same for all compositions. \cite{burrows07b} and \cite{guillot10} studied the effect of the presence of heavy-elements and different opacities. The \cite{burrows07b} work scaled abundances of elements up by various factors, recalculated equilibrium chemistry, and used the chemical abundances to calculate opacities. Although the gas opacities were tied to atmospheric metallicity, grain opacities were not considered. \cite{guillot10} used a semi-grey atmosphere model which is parameterized as a function of mean visible and thermal opacities. No attempt was made to tie the grain opacity to the atmospheric metallicity in a self-consistent manner. \cite{valencia13} modeled the low-mass dense planet GJ-1214b, composed mainly of \h2o. They examined the grains' impact on the inferred envelope composition by using a fit for gas opacity tables.  However they did not calculate the grain opacity directly, but used an extrapolation from the opacity tables of \cite{alfurg94} .

Adding high-Z material to a planet does not only change the density everywhere inside the planet, but it also changes the opacity. The precise value of the opacity change depends on the chemical composition \citep{fortney08}, as well as on the details of the grain size distribution.  This size distribution depends, in turn, on the interplay of processes such as sedimentation and coagulation of grains \citep{podolak03,movsh08} as well as the condensation and evaporation of volatiles at different depths within the planet. 
The numerical complexity of the problem makes it difficult to do a study of the relevant parameter space for giant planet evolution which includes all the details of the microphysics.  On the other hand, the high-Z material undoubtedly changes the opacity and can strongly influence the structure and evolution of the planet. 
In this study we use a relatively simple model to investigate the effect of the planetary composition and opacity on the long-term evolution. We investigate the effect of metallicity on the mass-radius relation for various planetary masses and compositions when the high-Z component is included in both the equation of state and opacity calculations. 

\section{Model}
\subsection{Planetary Evolution Code}
We compute the evolution of a non-rotating spherically symmetric protoplanet using a stellar evolution code that has been adapted to deal with bodies of planetary mass \citep{helled06,helled08a,vazanhel12}. The code uses an adaptive mass zoning that is designed to yield optimal resolution \citep{kovetz09}. The planet is assumed to be composed of a solar mix of hydrogen and helium with an additional high-Z component.  This additional component can be either \sio2 (rock) or \h2o (water), and can be either segregated in a central core or mixed with the hydrogen and helium. Models with both a core and a heavy element envelope are also considered.  Details of the code can be found in \cite{kovetz09}.

\subsection{Equations of State}
For hydrogen and helium we used the equation of state tables from the work of \cite{scvh}. For given pressure and temperature, these tables list---either for H or for He---the density $\rho(p,T)$, the specific entropy 
$s(p,T)$ and the specific internal energy $u(p,T)$. We have used our own stellar EOS \citep{kovetz09} in order to extend these tables to lower pressures and temperatures.  

For \h2o and \sio2 we compute an equation of state based on the quotidian equation of state (QEOS) described in \cite{qeos88}. This combines the Cowan ion equation of state with the Thomas-Fermi model for the electrons. 
Following Young and collaborators \citep{youngcor95, qeos88} we have constructed our own, more extensive, QEOS tables. Since stable molecules do not exist according to the Thomas-Fermi model, we treated  compounds, such as \h2o or \sio2, as mixtures of atoms, in which the individual Wigner-Seitz cells were fixed by requiring their surface pressures to be equal. Isotherms with segments along which $\partial p(\rho,T)/\partial\rho < 0$ were replaced by phase equilibria, determined by equating the specific Gibbs functions of the two phases. For given density $\rho_Z$ and temperature $T$, the QEOS tables yield the pressure $p_Z(\rho_z,T)$, the specific
internal energy $u_Z(\rho_z,T)$, and the specific entropy $s_Z(\rho_z,T)$; here Z denotes either \sio2 or \h2o.

\begin{figure}
\centerline{\includegraphics[angle=0, width=90mm]{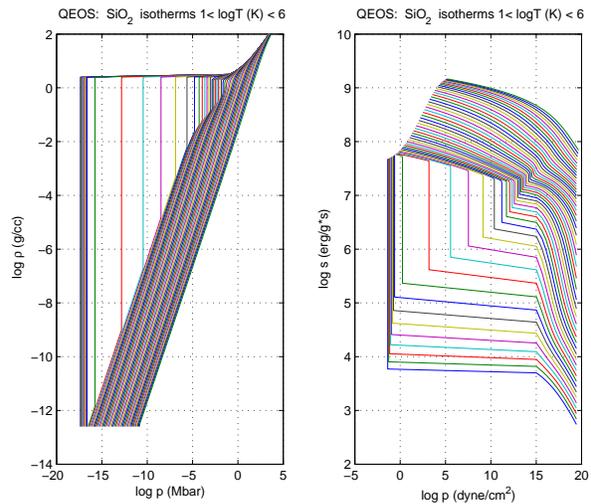}}
\caption{Equation of state of \sio2. Left panel - density vs. pressure for isotherms.
Right panel - entropy vs. pressure for isotherms.}\label{fqeoss}
\end{figure}

\begin{figure}
\centerline{\includegraphics[angle=0, width=90mm]{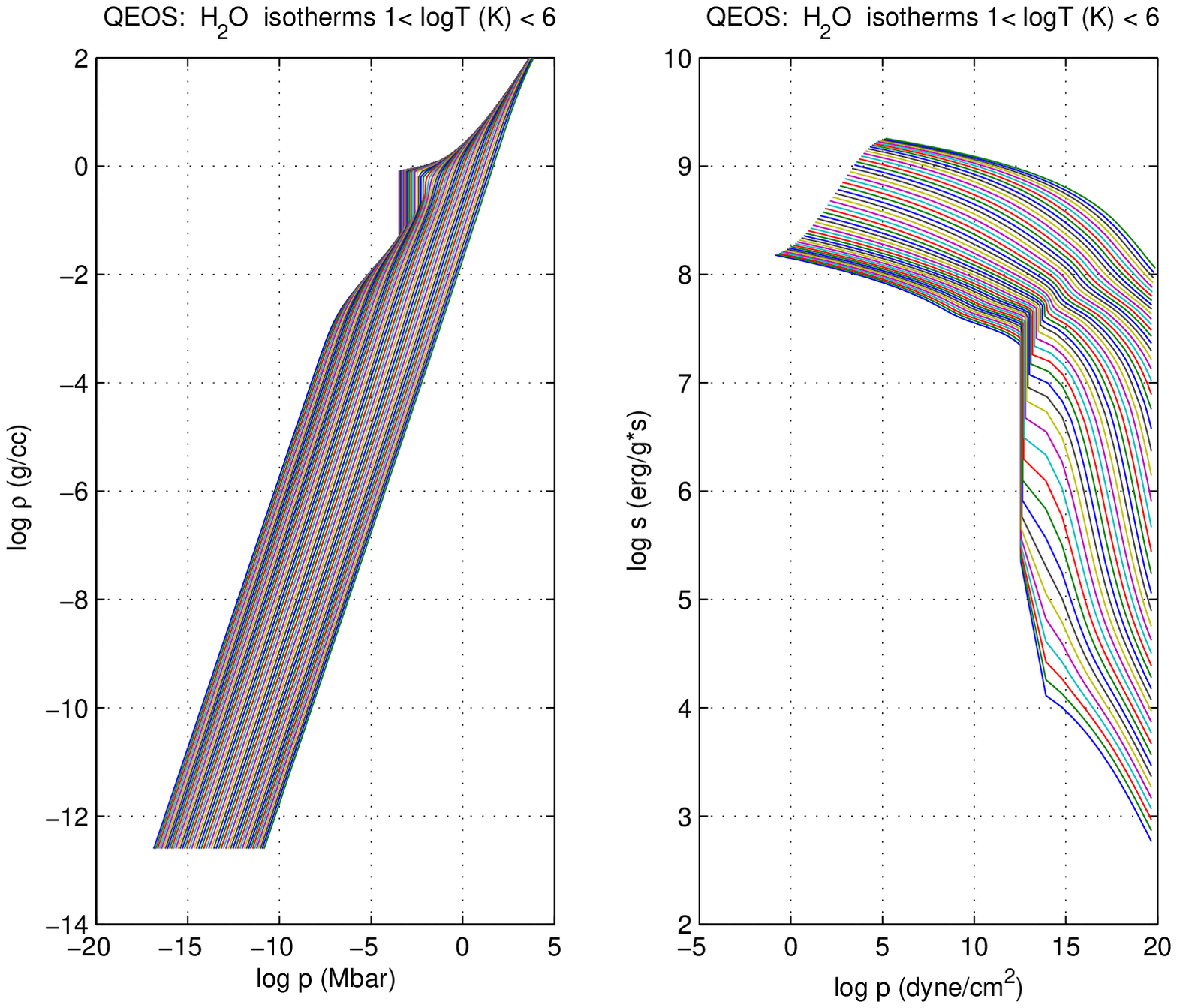}}
\caption{Same as Fig.\,\ref{fqeoss} but for \h2o. }\label{fqeosh}
\end{figure}

\begin{figure}
\centerline{\includegraphics[angle=0, width=84mm]{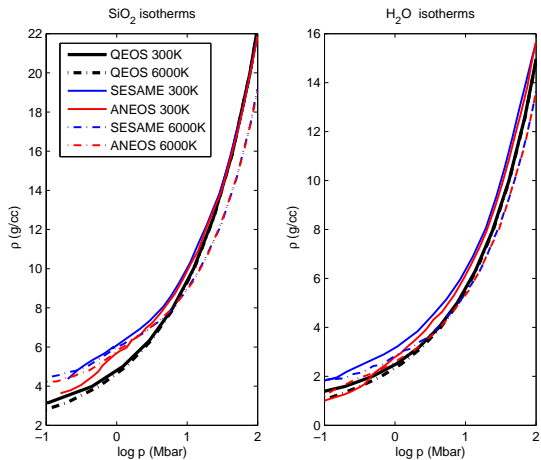}}
\caption{Isotherms of \sio2 (left) and \h2o (right) as computed from ANEOS (red), SESAME (blue) and the QEOS (black) for 300\,K (solid) and 6000\,K (dash-dot).}\label{qeoscomp}
\end{figure}

Figures \ref{fqeoss} and \ref{fqeosh} present the resulting QEOS for a large range of temperatures and densities for quartz (\sio2) and water (\h2o), respectively. Isotherms are presented for density-pressure (left panels) and entropy-pressure (right panels). The density jumps indicate a phase transition from vapor to solid. As expected, the density discontinuity disappears at higher temperatures. It is important to note that in addition to the obvious density change accompanying the phase change, there is also a decrease in entropy during a phase transition from vapor to solid. The jump in entropy due to a phase transition can have an important effect on the energy budget of the planet. 
Figure \,\ref{qeoscomp} compares the QEOS (black) with the ANEOS (red) and SESAME (blue) equations of state as given in \cite{baraffe08}.  
The two isotherms are for temperatures of 300 K (solid) and 6000 K (dashed-dot). As can be seen from the figure, there is a good agreement between the widely used EOS calculations and the QEOS. 

For the treatment of a mixture of hydrogen, helium and a heavy-Z compound---either quartz or water---we proceed as follows:
If $m$ is the mass of the mixture, then  $V=m/\rho\,$, where $\rho$ is its
density. Similarly, $N_iv_i=V_i=m_i/\rho_i\,$,  where $m_i$ is the mass of the $i$'th species,
and  $\rho_i(p,T)$ its density. The additive-volume law (Eq.\ref{mix8} of the Appendix) can be written in the form
\begin{equation}\label{mix9}
\frac{1}{\rho}=\sum {X_i\over\rho_i(p,T)},
\end{equation}
where $X_i=m_i/m$  is the mass fraction of the $i$'th species.
All the species are manifestly at the same
pressure and temperature, and the last equation can be regarded as the (implicit) formula
for the pressure of the mixture as a function of the density, the temperature and the
mass fractions.

For a mixture of hydrogen, helium and a compound Z---either SiO$_2$ or H$_2$O---we must
take account of the fact that the \cite{scvh} tables for H or He list $\rho_H(p,T)$ or
$\rho_{He}(p,T)$, whereas our quartz or water tables list $p_Z(\rho_Z,T)$. We therefore
write eq.\ref{mix9} in the form
\begin{equation}
{1\over\rho}={X\over\rho_H(p_Z(\rho_Z,T),T)}+{Y\over\rho_{He}(p_Z(\rho_Z,T),T)}+{Z\over\rho_Z},
\end{equation}
and, for given $(\rho,T,X,Y,Z)$, solve this equation by iterating on $\rho_Z$. When this
has converged, the three materials are all at the same pressure $p=p_Z$ and temperature $T$, and we have the three densities,
as well as the pressure, as functions of $(\rho,T,X,Y,Z)$. Moreover, we also have the specific
internal energies and entropies for each of the three species. See the Appendix for mixture calculation of energy and entropy.


\subsection{Opacity Calculation}\label{opcal}

In order to properly account for the high-Z material, we consider both grains and gas in the opacity calculation.  The grain opacity depends on the number of grains, their composition and size distribution.  The exact form of the size distribution depends on the details of the grain microphysics, and this is beyond the scope of our study.  For simplicity we consider a single size distribution of spherical grains.  Their scattering and absorption cross sections are computed using the standard formulas of Mie theory \citep[see, e.g.][]{hulst57}. The complex refractive index of \h2o is taken from \cite{warren84}, while for \sio2 we use the refractive index for olivine taken from \cite{dorschner95}. 

The cross section for extinction at a frequency, $\nu$ by a grain of radius $a$ is $\sigma_{\nu}=Q_{\nu}(a)\pi a^2$ where $Q_{\nu}(a)$ is the extinction efficiency given by Mie theory, calculated as described in \cite{podzuck04}. Once the efficiency is known, the opacity for grains of radius $a$ at this frequency can be computed from 
\begin{equation}\label{kappanu}
\kappa_{\nu}^{grain}=\frac{n(a)Q_{\nu}(a)\pi a^2}{\rho_{gas}}
\end{equation}
where $\rho_{gas}$ is the background gas density, and $n(a)$ is the number density of grains. 
Both the gas and grains contribute to the radiative opacity which is computed as the sum of gas and grain opacities, integrated over the frequency in order to form the Rosseland mean. The gas opacities are taken from tables of opacity kindly supplied by A. Burrows and based on the work of \cite{burrows07}. 

In order to approximate the vaporization of grains, we employ the following algorithm: at each point in the planetary envelope we compute the partial pressure of the high-Z component assuming that all of the material is in the gas phase.  We then compare this with the equilibrium vapor pressure of the material at that temperature.  If the partial pressure is lower, we assume that all the high-Z material is in the form of vapor, otherwise it is assumed to be in the form of grains.  Although this is a simplified approximation of the actual situation it is found to reproduce the results of more detailed calculations. The expressions for the vapor pressures of ice and rock as a function of temperature are taken from \cite{podolak88}. At very high densities, such as in the planetary core/center, the radiative opacity becomes very high, and the electron thermal conductivity becomes a much more efficient means of transferring heat. In this regime it is the conductive opacity that dominates. The conductive opacity is calculated using the tables presented by \cite{potekhin99}. The effective opacity is taken to be the harmonic mean of the opacities; $\kappa=1/({\kappa_{rad}}^{-1}+{\kappa_{cond}}^{-1})$ where $\kappa_{rad}$ and $\kappa_{cond}$ are the radiative and conductive opacities, respectively. \par

Planetary evolution calculations commonly use opacity tables based on the work of \cite{pollack85} for early solar nebula (hereafter - {\it solar opacity}) or of \cite{d'alessio01} for T Tauri disks.  These opacities are calculated for grains containing a mixture of materials.  In order to compare our opacity calculations more directly with those of Pollack and D'Alessio we compute the opacity for grains consisting of a 0.63:0.37 mixture of \h2o and \sio2 by mass.  This is approximately the ratio expected for a solar mix of elements. We assume that the materials are well mixed, and compute the refractive index of the mixture using the Clausius-Mossotti relation \citep{marion65}. 

\begin{figure}
\centerline{\includegraphics[angle=0, width=84mm]{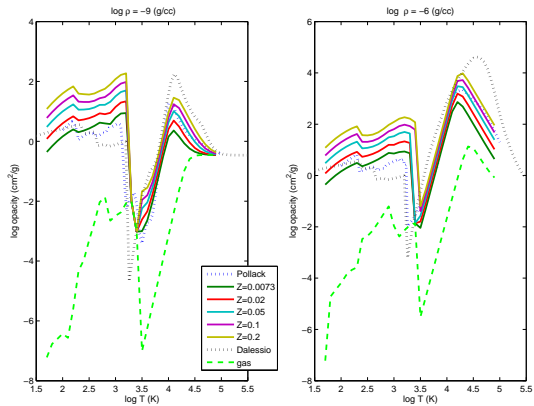}}  
\caption{Opacity as a function of temperature for $\log \rho=-9$ (left panel) and $\log \rho=-6$ (right panel) for a grain radius of $10^{-4}\,$cm composed of \h2o and \sio2. The solid curves are our calculation for different values of Z.  The opacities of \citealt{pollack85} (blue dashed curve) and \citealt{d'alessio01} (black dashed curve) are shown for comparison. The green dashed curve shows the opacity due to gas alone \citep{burrows07}.}
\label{opz-9}
\end{figure}

Figure \,\ref{opz-9} shows the calculated opacity (solid curves) as a function of temperature for gas densities of $10^{-9}$\,\gc3 (left panel) and $10^{-6}$\,\gc3 (right panel). Unless otherwise stated, we assume a grain size of $10^{-4}$\,cm for all calculations,  and that the heavy elements are mixed with the gas. We present here the opacity-temperature relation for Z$=0.0073, 0.02, 0.05, 0.1,$ and 0.2. For comparison, also plotted are the opacities calculated by \cite{pollack85} (blue dotted curve) and \cite{d'alessio01} (black dotted curve). Both the Pollack {\it et al.}~and the D'Alessio~{\it et al.} calculations are for solar composition, i.e. Z$\sim 0.02$. 
As expected, a larger abundance of heavy elements increases the opacity values. One can see the sharp drop around $\log T=2$ where the ice evaporates from grains and the sharper drop around $\log T=3.5$ where the rock evaporates. At higher temperatures the heavy elements affect the opacity as vapors. The rise in opacity due to the influence of the H$^-$ ion can be seen. It starts right after the grains evaporate, then as the temperature increases a subsequent decrease occurs as the Kramers-type opacity becomes important. This result fits well with the Pollack {\it et al.} and D'Alessio {\it et al.} tables. At the higher density, the ionization occurs at higher temperatures, and the asymptotic regime is not reached in the temperature range we consider. The gas opacity (green dashed curve) is also plotted to emphasize the importance of the grain contribution. 

\begin{figure}
\centerline{\includegraphics[angle=0, width=84mm]{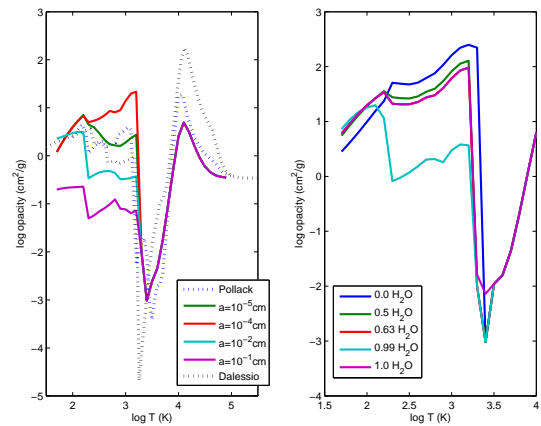}}
\caption{Left panel - Opacity as a function of temperature for different grain sizes for the case where $\log\rho=-9$, Z$=0.02$. Also shown are the opacities of \citealt{pollack85} (blue dashed cuvre) and \citealt{d'alessio01} (black dashed curve) for comparison.  Right panel - Opacity as a function of temperature for grains with different ice fractions. In all cases Z$=0.1$, $\log\rho=-9$ and the grain size is $10^{-4}\,$cm.}
\label{opg-9}
\end{figure}

The grain size distribution in planetary interiors is unknown and, in addition, the grain radii can change with time due to physical processes such as coagulation, sedimentation, evaporation, etc. The assumed grain radius has a significant effect on the calculated grain opacity by changing the surface area that corresponds to a given mass. While for a given composition small grains have larger surface area, if they are small enough their extinction efficiency drops quickly with radius. The left panel of Figure \,\ref{opg-9} shows the effect of varying the grain size between $10^{-5}$ and $10^{-1}$\,cm  for Z$=0.02$ and constant component (\h2o/\sio2 ratio).  At the lowest temperatures, where the longest wavelengths contribute to the Rosseland mean, the opacity drops by roughly an order of magnitude as the grain radius increases from $10^{-2}$ to $10^{-1}$\,cm. As the temperature increases and shorter wavelengths contribute to the Rosseland mean, smaller grains yield larger opacities until the temperature of vaporization of the rock component is reached.  At this point the opacity becomes independent of grain size.  Clearly, the grain size has a substantial effect on the opacity, and we can therefore expect the planetary evolution to depend, not only on the composition, but also on the assumed grain sizes. These effects are presented in detail in the following sections.\par

Finally, there is the dependence on composition.  First of all, different materials have different refractive indexes, but more importantly, they evaporate at different temperatures. Since the vaporization temperature of \h2o is much lower than that of \sio2 the drop in opacity at the ice vaporization point will vary as the \h2o/\sio2 ratio in the grain varies.  The right panel in Figure \ref{opg-9} shows the opacity as a function of temperature for grains with different mass fractions of ice for the case of $10^{-4}$\,cm grains at a gas density of $10^{-9}$\,\gc3.  In all cases (except for that of no ice), the drop in opacity at the ice evaporation point around $\log T=2.3$ is clearly visible.  Note that there is a considerable difference between a grain with an \h2o mass fraction of 0.99 and a grain with an \h2o mass fraction of 1.0.  This is due to the residual \sio2 in the grain which is absent in the pure \h2o case. 
After the grains evaporate the different curves converge since the gas opacity calculation does not account for the composition of the high-Z material but only for its mass fraction. The opacity calculation after grain evaporation appears in the Appendix.

The ratio of gas to grain opacity varies with grain size, Z-fraction, density and temperature. Grains can increase the opacity by 2-8 orders of magnitude. For low temperatures the gas opacity is negligible in comparison to that of the grains. After the grains evaporate, the contribution of the grains' vapor to the opacity is 1-3 orders of magnitude. 
Therefore, even a small fraction of grains in the planetary envelope can decrease the radiative flux substantially, and slow the contraction of the planet.


\section{Modeling the Planetary Evolution}

\subsection{Evolution of Various Compositions - 1\mj with Solar Opacity}

\begin{figure}
\centerline{\includegraphics[angle=0, width=84mm]{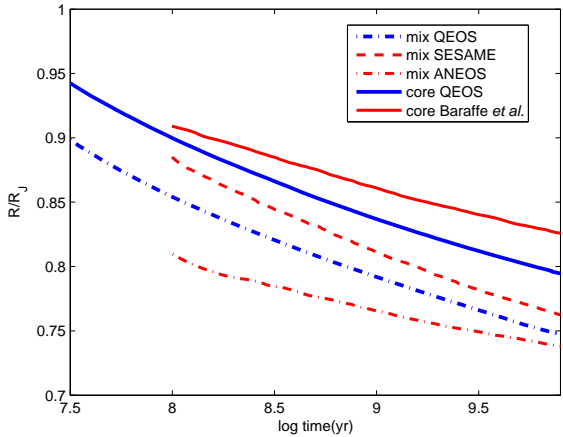}}
\caption{Radius as a function of time for a 1\mj planet with a solar ratio of hydrogen to helium and $Z=0.5$ in the form of \h2o. The blue solid curve is for a model where all the \h2o is in the core. The blue dash-dot curve shows the evolution for the case where the \h2o is mixed with the H/He. Shown for comparison are the models of \citealt{baraffe08} for \h2o core (red solid) and for the mixture using the ANEOS (red dash-dot) and SESAME (red dash) EOS.}\label{brf6b}
\end{figure}

In order to understand the effect of the composition and its depth dependence on the planetary evolution we first calculate the evolution of a 1\mj planet with various compositions and internal structures. Figure \ref{brf6b} shows the planetary radius as a function of time for a 1\mj planet composed of 0.5 \h2o by mass and H/He in the solar ratio. For comparison, also presented are the results found by \cite{baraffe08}. The dashed curves show evolutionary curves for the case where the \h2o and the H/He are mixed throughout the planet. The blue dash-dot curve is for our calculation with QEOS, and the red curves are the calculations of \cite{baraffe08} for the ANEOS EOS (dash-dot curve) and the SESAME EOS (dashed curve). As can be seen from the figure, the QEOS gives an evolutionary track that falls between that of ANEOS and SESAME. The solid curves are for the case where all of the \h2o is sequestered in the core, and the envelope is a pure H/He mix.  The blue curve is for our calculation with QEOS, the red curve is for the calculation of \cite{baraffe08} using the ANEOS EOS for the core.  It should be noted that in addition to the differences in the equations of state, the conductive opacities associated with ANEOS also differ somewhat from the \cite{potekhin99} opacities used in our code. In spite of these differences, the agreement between the evolution curves is excellent.

The difference between the models due to the uncertainties in the EOS amount to approximately 10\% of the radius as found by \cite{baraffe08}. For the case of the particularly high value of Z = 0.5, presented in figure \ref{brf6b}, the distribution of the material is important. The mixed and core-envelope (metal-free) models differ in radius by 5\%. Replacing \h2o by \sio2 in core-envelope model leads to a $\sim$ 5\% smaller radius, while in the totally mixed model the difference is $\sim$ 15\%. 

\begin{figure}
\centerline{\includegraphics[angle=0, width=84mm]{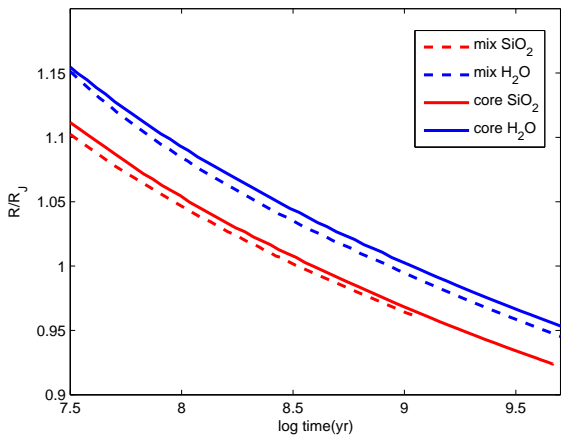}}
\caption{Radius as a function of time for a 1\mj planet with $Z=0.2$. The blue and red curves are for \h2o and \sio2, respectively.  Solid curves correspond to cases in which all the high-Z material is in the core, while dashed curves correspond to fully mixed planets.}\label{brf6c}
\end{figure}

For lower values of Z the models become less sensitive to the distribution of the high-Z material. Figure \ref{brf6c} shows the evolutionary tracks for models that have $Z=0.2$. The blue curves correspond for \h2o while the red curves correspond for \sio2. The solid curves represent evolution models assuming that all the heavy elements are concentrated near the center with a metal-free H/He envelope, while the dashed curves are the results for the cases in which the heavy element mass is homogeneously  mixed within the planet. For this particular case, the difference between \h2o and \sio2 models is $\sim$ 5\% while the difference between models with a core and fully mixed models is negligible. We can therefore conclude that for giant planets with relatively small factions of high-Z material (Z$\lesssim 0.2$) no inner structure information can be derived from the mass-radius relation. \par 


\subsection{Modeling the Planetary Evolution - 1\mj with Self-Consistent Opacity}

The models presented above were computed using standard opacity tables assuming a solar-composition that are common in planetary evolution modeling ({\it solar opacity}). However, as mentioned above, the additional high-Z material contributes to the opacity as well. If the material is in the form of grains, this contribution becomes substantial. As a result, we repeat the above calculations with our own grain+gas opacities.  It is known that the grain size distribution inside an evolving planet is different from that observed in the interstellar environment, since the grains in a planet can grow by coagulation \citep{podolak03,movsh08,movsh10}.  Here we only consider a single grain size distribution, in order to estimate how important opacity effects are. 

\begin{figure}
\centerline{\includegraphics[angle=0, width=84mm]{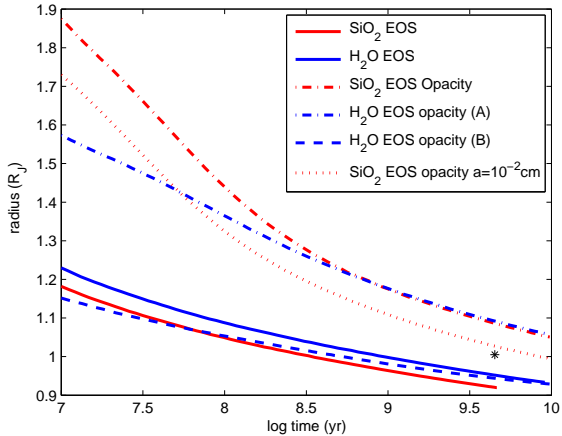}}
\caption{Radius as a function of time for a 1\mj planet with $Z=0.2$.  Half the high-Z material is in the core (0.1\mj) and the rest is homogeneously mixed throughout the envelope.  If we use solar opacity, the evolution is given by the solid curves for \h2o (blue) and \sio2 (red).  The blue dashed curve is for our opacities assuming 10$^{-4}\,$cm grains and pure \h2o. The dash-dot curves are for \h2o grains with 1\% \sio2 included (blue) and for pure \sio2 (red).  The dotted curve is for \sio2 with  10$^{-2}\,$cm grains.  See text for a full discussion.}\label{Rop1}
\end{figure}

Figure \ref{Rop1} shows the evolutionary track for a 1\mj planet with $Z=0.2$.  For these models half the high-Z material is in the core (core mass is $\sim$30 \me) while the other half is uniformly mixed throughout the envelope. It will be recalled that for $Z=0.2$ the radius is not sensitive to the exact distribution of the high-Z material, and we have chosen something intermediate between having all the high-Z material in the core and having it mixed completely throughout the planet. 
The solid curves show the evolution for the case with solar opacity as presented above, for \h2o (blue) and \sio2 (red). 
The red dash-dot curve shows the \sio2 model with the grain opacity of the additional material included in the form of $10^{-4}$\,cm-sized grains.  The \sio2 track shows that this additional opacity can have a much larger effect on the radius than uncertainties in either the composition or the equation of state.  Increasing the grain size to $10^{-2}$\,cm, reduces the total surface area of the grain component, and this reduces the opacity accordingly, as shown by the red dotted curve. Clearly, larger grain sizes result in lower opacity which accelerates the planetary contraction, and leads to more compact configurations at a given age.  {\it We conclude that the opacity of the high-Z material has a substantial effect on the planetary  evolution, and that the effect of different grain sizes is more significant than the effect of composition.} Therefore, one must be cautious when using the mass-radius relationship to infer the planetary composition. \par

The case with pure \h2o opacities, corresponding to $Z=0.1$ in the envelope (blue dash curve), gives an evolutionary track much more similar to those computed with the solar opacities \citep{pollack85} . This is because the \h2o grains evaporate at $T\sim 200$\,K. The \cite{pollack85} opacities contain refractory materials in addition to ice, and therefore in that calculation, grains survive and contribute to the opacity at much higher temperatures. The sensitivity to even a small amount of refractory material is shown by the red dash-dot curve, where the ice grains are now composed of 99\% \h2o and 1\% \sio2.  In this case, the planetary radius is again much larger than that computed using the solar opacities. Even a very small fraction of silicate grains is enough to affect the opacity, and therefore the planetary contraction, significantly. The merging of the \h2o and \sio2 curves at the later stages of evolution appears to be coincidental. Jupiter is represented by the black asterisk. 
These results demonstrate the importance of opacity on planetary contraction, and emphasize the need to model this effect properly. Clearly the impact of different grain sizes and their corresponding opacities on the planetary evolution is greater than the effect of the assumed EOS, the planetary composition, and the internal structure. \par

\begin{figure}
\centerline{\includegraphics[angle=0, width=84mm]{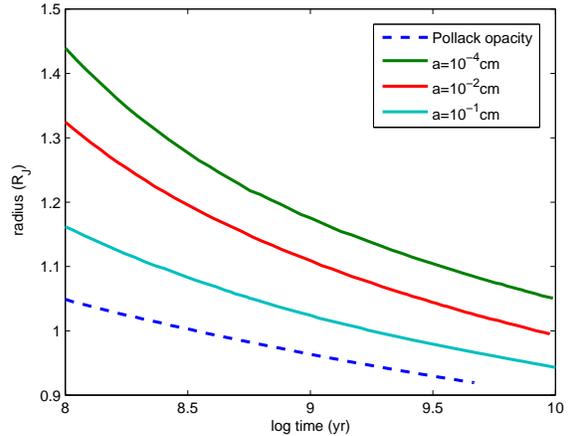}}
\caption{Same model as in figure \ref{Rop1}, for \sio2 grains with radii of 10$^{-4}\,$cm (green), 10$^{-2}\,$cm (red), and 10$^{-1}\,$cm (cyan).  The evolution for the solar opacity is also shown (blue dashes).}\label{Rgrain}
\end{figure}

As discussed earlier, the assumed grain size in the opacity calculation has a substantial effect on the planetary evolution. We next repeat the evolution calculation of the previous structure (0.1\mj core + envelope with $Z=0.1$), for various grain sizes. The heavy elements in these simulations are represented by \sio2. We also show an evolutionary track without the additional high-Z opacity for comparison. Here, different grain sizes appear to change the planet radius by 15\%. The results are shown in figure \ref{Rgrain}. \par 

\begin{figure}
\centerline{\includegraphics[angle=0, width=84mm]{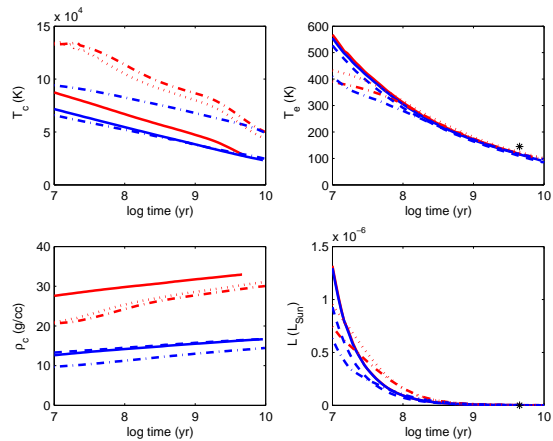}}
\caption{ Central temperature (upper-left), central density (lower-left), effective temperature (upper-right) and luminosity (lower-left) as a function of time for a 1\mj planet, for the cases of Fig.\,\ref{Rop1}. Again, the blue and red curved correspond to \h2o and \sio2, respectively, and the solid curves represent cases with solar opacity. Black asterisks correspond to Jupiter.}\label{trol}
\end{figure}

Figure \ref{trol} shows the evolution of the physical properties, i.e., the central and effective temperatures, central density and luminosity, of the cases presented in figure \ref{Rop1}.
Central temperature (upper-left panel) of the planets with the opacity calculation included are higher during evolution (slower heat release), and the central density (lower-left panel) are lower, respectively. It can be seen from the figure that planets with \sio2 cores cool at a faster rate, probably due to more efficient conductivity. As expected, the central density of planets with \sio2 cores are higher than those of \h2o cores. From the panel which shows the effective temperature we can conclude that during the early phases of the planetary evolution the differences between the different models are significant but they decrease with time, and after $\sim 10^8$ years the differences are relatively small. When fitting Jupiter's observed temperature (black asterisk) we find that all the models we consider fit equally well, and we therefore suggest that without additional constraints on the density profile such as gravity data, which is the case for extrasolar planets, it is not possible to distinguish between the different configurations just by measuring the mass, radius, and effective temperature. \par


\subsection{Modeling the Planetary Evolution - 1\mj with Irradiation} 

Another effect that can affect the radius of an evolving planet is irradiation by the parent star. We investigate this effect by assuming that the irradiation is incident on the planet isotropically. Details on the implementation of irradiation in the model can be found in the Appendix.  
We run models where the irradiation is equivalent to that received by a planet at 1\,AU, 0.1\,AU, and 0.05\,AU from a sun-like star, and compare these cases to the case of an isolated planet (no irradiation).  

\begin{figure}
\centerline{\includegraphics[angle=0, width=84mm]{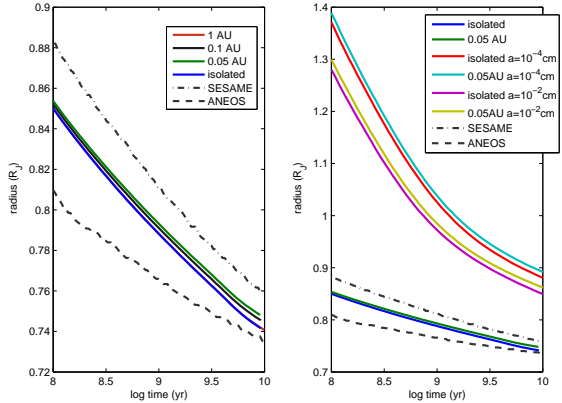}}
\caption{Left panel - Radius as a function of time for a 1\mj planet with $Z=0.5$ \h2o mixed throughout the volume.  Irradiation is from an external source at distances of 1\,AU (blue), 0.1\,AU (green), and 0.05\,AU (red) compared to an isolated planet (cyan).  These are compared with the calculations of \citealt{baraffe08} using the ANEOS (dotted curve) and SESAME (dash-dot) EOS.   Right panel - The effect of including the grain opacity of all the high-Z material.  Shown are comparisons between an isolated planet with 10$^{-4}\,$cm grains (red) compared to an irradiated planet at 0.05\,AU (cyan) and an isolated planet with 10$^{-2}\,$cm grains (purple) compared to an irradiated planet at 0.05\,AU.  The models of the left hand panel are included for comparison.}\label{irrad1}
\end{figure}

The results are shown in the left panel of Figure \ref{irrad1}. In order to compare the effect of irradiation with that of the assumed EOS, also presented are the evolutions with the ANEOS and SESAME equations of state as computed by \cite{baraffe08}. 
As can be seen from the figure, the effect of irradiation on the planetary radius as a function of time is much smaller than the uncertainties in the equation of state. 
As expected, for planets with significant abundances of heavy elements the effect of irradiation on the planetary radius at later times is small, even in comparison to the EOS uncertainty. Also shown in the right hand panel of Figure \ref{irrad1} are evolutions in which the opacity effect is included. We consider both the isolated and irradiated cases, and two different grain sizes. 
Although irradiation increases the temperature of the outer layers, the temperature at the photosphere does not reach values large enough to evaporate all the grains, even at 0.05\,AU. As a result, the opacity remains relatively high and leads to a slower contraction than for the case with the solar opacity. Clearly, the change in radius due to irradiation is small compared to the increase due to the inclusion of the opacity of the additional high-Z material. It should be noted that in computing the effect of irradiation we considered only the incident radiation from the star. If ohmic dissipation operates in the convective zones of close-in planets \citep{batstv10}, or occurs in the atmospheres of these planets \citep{perna10}, the effect of irradiation on radius could be larger than anticipated.

\subsection{Modeling the Planetary Evolution - 20\me}

\begin{figure}
\centerline{\includegraphics[angle=0, width=84mm]{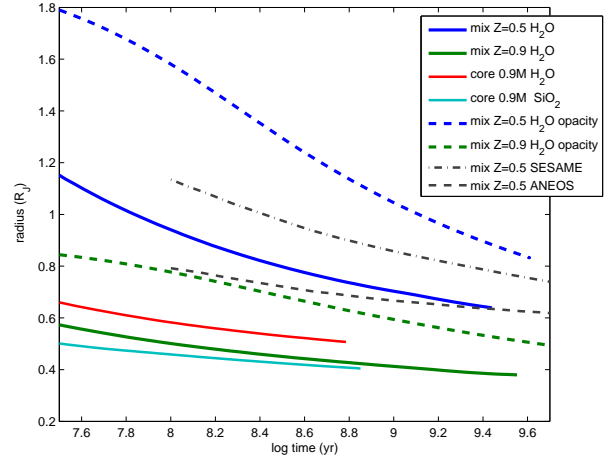}}
\caption{Radius as a function of time for a 20\me planet.  Shown are models for \h2o planets where the \h2o is mixed throughout the planet for $Z=0.5$ (blue) and $Z=0.9$ (green) - Solid curves present evolution using solar opacity and dashed for the same model with our opacity calculation for appropriate Z fraction. Also shown are two examples with $Z=0.9$ when all the heavy material segregated in the core for \h2o (red) and \sio2 (cyan). For comparison we show models of \citealt{baraffe08} where the \h2o, with $Z=0.5$, is completely mixed throughout the planet, and the EOS is computed from ANEOS (dot grey) and SESAME (dash-dot grey).} \label{Rop20me}
\end{figure}

In this section we model the evolution of planets with masses of 20\me, representative of Uranus and Neptune, and intermediate-mass extrasolar planets.
Figure \ref{Rop20me} shows some representative cases.  The blue solid curve shows the evolutionary track for a planet containing $Z=0.5$ of \h2o that is completely mixed throughout the volume. For comparison, also shown are the models computed by \cite{baraffe08} using the ANEOS (dots) or SESAME (dash-dot) equations of state. 
Increasing the heavy element fraction in the planet up to $Z=0.9$ (green curve) decrease the radius by about 70-80\%. In that case, which might be a representative case for intermediate-mass planets, the actual distribution of the heavy elements can lead to differences of about 15\% in the planetary radius (red curve). The change in radius due to the composition of the heavy elements is significantly smaller (cyan). \par

Inserting self-consistent opacity calculation to the evolution of 20\me planets leads to a significant change in the planetary radius. When the heavy elements are assumed to be homogeneously mixed for cases with $Z=0.5$ (blue dashed) and $Z=0.9$ (green dashed) the radius increases by 50-70\% and 40-50\%, respectively. This is significantly larger than the change for the 1\mj case, where similar heavy element conditions (EOS and opacity) results in 20-30\% increase in radius.  It should be noted that for large fractions of heavy elements the derived evolutions are similar since under those conditions the bulk of the planet is convective which results in a similar efficiency of the heat transfer.  Indeed, \cite{valencia13} found that for the small dense planet GJ-1214b opacity doesn't change the evolution substantially. Their result is in agreement with our expectations for such an object.  For less dense objects, the presence of grains will have a much larger effect on the evolution. \par

\begin{figure}
\centerline{\includegraphics[angle=0, width=84mm]{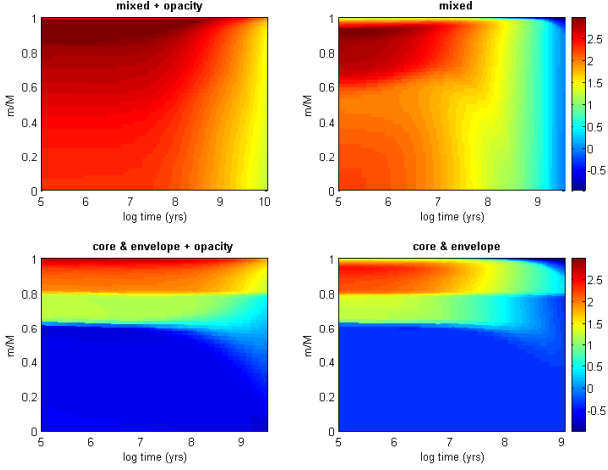}}
\caption{Planetary opacity within the planet as a function of time. We compare two cases of 20\me planets with $Z=0.8$: homogeneously mixed (upper panels) and core+envelope (lower panels), with solar opacity (left panels) and self-consistent opacity (right panels).}\label{Rop3d}
\end{figure}

Figure \ref{Rop3d} shows the opacity distribution within the planet as time evolves with (left panel) and without (right panel) self-consistent opacity calculation.  Two cases are presented: mixed planets (upper panel) in which the heavy elements are homogeneously distributed and internal structure of a core and envelope (lower panel). In both cases the heavy element mass fraction is taken to be $0.8$. As can be seen from the figure, the region which is mostly affected by the radiative opacity, in both cases, is the outermost layers in which the temperatures are lowest and grains can survive. For the core and envelope models the conductive opacity is more dominant throughout the core region, and therefore the opacity is low. The opacity in the outer layers, however, is high, and as a result, the contraction of the planet is slower.

\begin{figure}
\centerline{\includegraphics[angle=0, width=84mm]{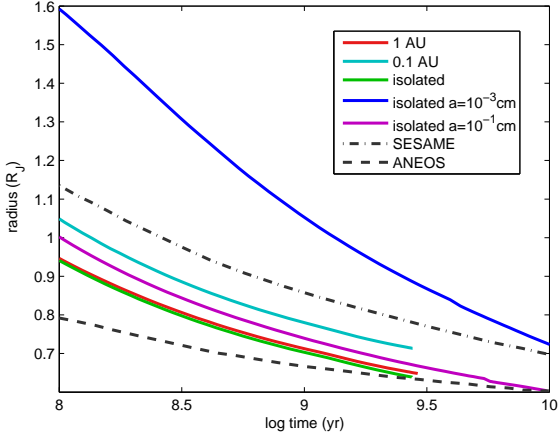}}
\caption{Similar to Figure \ref{irrad1} (right panel) but for 20 \me planets. Shown are comparisons between an isolated planet with 
solar opacity (green) compared to an irradiated planet at 0.1\,AU (cyan) and 1\,AU (red) and isolated planets with our opacity calculation for $10^{-3}$cm grains (purple) and 10$^{-1}$cm grains (blue) .}\label{irr20me}
\end{figure}

The effect of irradiation on the evolution of 20 \me planet is shown in Figure \ref{irr20me}.  Similarly to the case of 1 M$_J$ planet, the change in radius due to the opacity is larger than the effects of irradiation and assumed EOS. However, irradiation has a greater impact on 20 \me planets. The planetary radius can change by up to 15\%. Again, it is found that the uncertainty in grain size leads to a significant difference in the derived radius-mass relation, of the order of 50\%. 
Here, too, we can conclude that the mass-radius relationship does not necessarily yields the planetary composition and internal structure.\par


\section{Summary}
We have computed evolutionary tracks for planets of 1 M$_J$ and 20 M$_{\oplus}$. We have explored the effects of composition (\h2o vs. \sio2), high-Z-material distribution (core-envelope vs. uniformly mixed), grain opacity, and irradiation. We find that the most important effect is that of the grain opacity due to the additional high-Z material in the envelope.  This has the potential of increasing the computed radius of the planet by several tenths.  However, the precise effect is difficult to compute accurately.  Our calculation assumes that the grains are affected by temperature changes during evolution, but that the ratio of high-Z material to H/He remains constant with time.  Coagulation and sedimentation of grains will change this ratio as the planet evolves, and we may expect that the additional grain opacity will decrease with time.  We hope to address this complex problem in a future study. \par

\section*{Acknowledgments}
We thank D. A. Young for sharing with us the QEOS tables and A. Burrows for the gas opacity tables. We also acknowledge the Israel Science Foundation grant \# 1231/10. 
AV acknowledges support from the Israeli Ministry of Science via the Ilan Ramon fellowship.

\section{Appendix}

\subsection{Mixtures}

For a collection of {\it pure, unmixed} substances, each of $N_i$ particles, the (extensive) 
Gibbs function is 
\begin{equation}\label{mix1}
G(p,T,N_1,N_2,...)=\sum N_i\mu_i(p,T),
\end{equation}
where $\mu_i(p,T)$ is the chemical potential of the $i$'th substance. We have
\begin{equation}\label{mix2}
\mu_i(p,T)=u_i-Ts_i+pv_i,
\end{equation}
where $u_i$, $s_i$ and $v_i$ are per particle. The volume of the $i$'th substance is $N_iv_i\,$. We have
\begin{equation}\label{mix3}
d\mu_i(p,T)=-s_idT+v_idp.
\end{equation}
Upon mixing, the (extensive) Gibbs function is of the same form as before, $G^M=\sum N_i\mu^M_i$, but with different $\mu^M_i$'s, which we expect to
depend, in addition to $p$ and $T$, on {\it ratios} of the $N_i$'s, or on the {\it concentrations} $c_i=N_i/N$,
where $N=\sum N_i$. Lewis and
Randall's formulae for an {\it ideal} mixture, which actually hold when the substances are ideal gases, are
\begin{equation}\label{mix4}
\mu^M_i(p,T,N_1/N,N_2/N,...)=\mu_i(p,T)+kT\ln {N_i\over N},
\end{equation}
whatever the nature of the substances, that is, whatever the $\mu_i(p,T)$ ($k$ denotes Boltzmann's constant).
According to this assumption, the Gibbs function for the ideal mixture is
\begin{equation}\label{mix5}
G^M=\sum N_i\mu^M_i=\sum N_i\mu_i(p,T)+kT\sum N_i\ln {N_i\over N}.
\end{equation}
We shall denote
\begin{equation}\label{mix6}
S_{mix}=-k\sum N_i\ln {N_i\over N},
\end{equation}
and justify this name later. From eq.\ref{mix5} we obtain
\begin{equation}\label{mix7}
dG^M=\sum N_i(-s_idT+v_idp)-S_{mix}dT-TdS_{mix}+\sum \mu_idN_i,
\end{equation}
from which we can read off
\begin{equation}\label{mix8}
V={\partial G^M\over\partial p}=\sum N_iv_i(p,T),
\end{equation}
which is the additive-volume formula. Secondly,
\begin{equation}\label{mix10}
S=-{\partial G^M\over\partial T}=\sum N_is_i(p,T)+S_{mix},
\end{equation}
which justifies the name `mixing entropy' (and the notation of eq.\ref{mix6}).
Finally,
\begin{eqnarray}
U&=&G^M+TS-pV\\
 &=&\sum N_i\mu_i-TS_{mix}+T(\sum N_is_i+S_{mix})-p\sum N_iv_i\nonumber
\end{eqnarray}
which, in accordance with eq.\ref{mix2}, reduces to the additive-energy formula
\begin{equation}\label{mix11}
U=\sum N_iu_i(p,T).
\end{equation}
It should, perhaps, be noted that the formula \ref{mix6} for the mixing entropy, the additive-volume
formula \ref{mix8}, and the additive-energy formula \ref{mix11} are not independent assumptions: they are
all consequences of the single Lewis-Randall assumption \ref{mix4}.

Consider now the differential of $U$, which is needed in a statement of the first law of thermodynamics:
\begin{equation}
dU=\sum (u_idN_i+N_idu_i)=\sum u_idN_i+\sum N_i(Tds_i-pdv_i).
\end{equation}
After some reduction ($N_ids_i=d(N_is_i)-s_idN_i\,$, $N_idv_i=d(N_iv_i)-v_idN_i$), this leads to
\begin{equation}
dU=\sum (u_i-Ts_i+pv_i)dN_i+T\sum d(N_is_i) -pdV,
\end{equation}
or, finally,
\begin{equation}\label{mix12}
dU+pdV=Td(\sum N_is_i)+\sum \mu_idN_i.
\end{equation}
Note that the mixing entropy does not appear in this formula. Nor do the Lewis-Randall chemical
potentials of eq.\ref{mix4}.
It is, of course, possible to use the differential of eq.\ref{mix6}, namely
\begin{equation}
dS_{mix}=-k\sum \ln\big({N_i\over N}\big)dN_i 
\end{equation}
in order to write eq.\ref{mix12} in the {\it alternative} form
\begin{equation}\label{mix13}
dU+pdV=Td(\sum N_is_i+S_{mix})+\sum \mu^M_idN_i,
\end{equation}
in which $S_{mix}$ and $\mu^M_i$ {\it do} appear, but the content of this equation is the same
as that of eq.\ref{mix12}. 

Evolutionary codes are based on the energy balance equation (the first law of thermodynamics)
\begin{equation}
{\partial u\over\partial t}+p{\partial\over\partial t}{1\over\rho}=q-{\partial L\over\partial m},
\end{equation}
where $u$ is the specific energy (energy per unit mass), $1/\rho$ is the specific volume,
$q$ is the rate of energy deposition per unit mass (in a planet, this may be due to frictional
heating of falling planetesimals), and $L(m,t)$ is the energy flux. The left hand side can be calculated
{\it either} from eq.\ref{mix12}, {\it or} from its equivalent eq.\ref{mix13}. In the first case, the mixing entropy
{\it need not be calculated\/}!

\subsection{Vapor Opacity Calculation}
In the regime where all the grains have evaporated, we used a simple analytical approximation based on the material on the website 'http://www.astro.princeton.edu/$\sim$gk/A403/opac.pdf'.  At the lower end of this temperature range the main contribution is from molecular opacity, which, for a mass fraction $Z$ can be approximated by
\begin{equation}\label{kmol}
\kappa_m\approx 0.1Z
\end{equation} 
At higher temperatures, $4\times 10^3\lesssim T \lesssim 8\times 10^3$\,K, the opacity due to the negative hydrogen ion, H$^-$, dominates. It can be approximated by
\begin{equation}\label{h-}
\kappa_{H^-}\approx 1.1\times 10^{-25}Z^{0.5}\rho^{0.5}T^{7.7}
\end{equation}
At temperatures where the material is partly ionized, (i.e $T\gtrsim 10^4$\,K) the opacity due to free-free, bound-free, and bound-bound transitions is well approximated by the {\it Kramers formula}
\begin{equation}\label{kramers}
\kappa_K=4\times 10^{25}(1+X)(Z+0.001)\frac{\rho_{gas}}{T^{3.5}}
\end{equation}
while for free electron scattering, a good fit to detailed theoretical calculations by \cite{buchler76} is given by: 
\begin{equation} \label{elecscat}
\kappa_e=0.2(1+X)\left[1+2.7\times 10^{11}\frac{\rho}{T^{2}}\right]^{-1}\left[1+\left(\frac{T}{4.5\times 10^8}\right)^{0.86}\right]^{-1}
\end{equation} 
The total radiative opacity can then be found from 
\begin{equation}
\kappa_{rad}\approx \kappa_m+({\kappa_{H^{-}}}^{-1}+(\kappa_e+\kappa_K)^{-1})^{-1}
\end{equation}
These factors can be combined into a formula that is valid for a large range of temperatures, $1.5\times 10^3 \leq T \leq 10^9\,$.

\subsection{Irradiation}
The following discussion is based on \cite{kovetz88}. Assuming a constant {\it net} outward flux $F\,$, the temperature distribution in a gray, plane-parallel atmosphere,
in the lowest approximation, is given by
\begin{equation}\label{bc3}
\sigma T^4(\tau)=(\threequarters\tau+\onequarter)F,   
\end{equation}
The constant $\onequarter$ is chosen so that the coefficient of $F$ is unity for $\tau=1\,$: thus, at the photosphere, where the optical depth is $\tau_S=1\,$ the actual temperature is equal 
to the effective temperature.

For a planet irradiated (normally) by a flux $\sigma T_{irr}^4\,$, the temperature distribution in the outer layers, assuming a {\it zero}
net outward flux, is given by
\begin{equation}\label{bc4}
\sigma T^4(\tau)=g(\tau) T_{irr}^4,
\end{equation}
where
\begin{equation}\label{bc5}
g(\tau)=\frac{3}{2}(1-\onehalf e^{-\tau}).
\end{equation}
Since the radiative transfer equation is linear, the two solutions can be superposed to yield the
temperature distribution for an irradiated planet in which the net outward flux is constant:
\begin{equation}\label{bc6}
\sigma T^4(\tau)=(\threequarters\tau+\onequarter)F+g(\tau)T_{irr}^4.
\end{equation}
Thus, at the photosphere, where $\tau=\tau_S=1\,$, the luminosity is
\begin{equation}\label{bc7}
L = 4\pi R^2F = 4\pi R^2\sigma [T^4-g(\tau_S) T_{irr}^4] .
\end{equation}
Note that, when $T_{irr} > 0\,$, the actual photospheric temperature $T$ is no longer equal to the effective temperature $T_E$.
In fact, according to the last equation,
\begin{eqnarray}\label{bc8}
T^4&=&T_{E}^4+g(\tau_S)T_{irr}^4\nonumber\\
   &=&T_{E}^4+1.224T_{irr}^4 .
\end{eqnarray}

\bibliographystyle{mn2e} 
\bibliography{allona}   

\end{document}